\documentclass[conference]{IEEEtran}
\IEEEoverridecommandlockouts
\usepackage{cite}
\usepackage{amsmath,amssymb,amsfonts}
\usepackage{algorithmic}
\usepackage{graphicx}
\usepackage{textcomp}
\usepackage{xcolor}
\def\BibTeX{{\rm B\kern-.05em{\sc i\kern-.025em b}\kern-.08em
    T\kern-.1667em\lower.7ex\hbox{E}\kern-.125emX}}

\title{An Empirical Evaluation of a Weighted PALU Model for Internet Traffic Networks\\
\thanks{Research was sponsored by the Department of the Air Force Artificial Intelligence Accelerator and was accomplished under Cooperative Agreement Number FA8750-19-2-1000. The views and conclusions contained in this document are those of the authors and should not be interpreted as representing the official policies, either expressed or implied, of the Department of the Air Force or the U.S. Government. The U.S. Government is authorized to reproduce and distribute reprints for Government purposes notwithstanding any copyright notation herein.}
}
    
\begin{document}

\author{\IEEEauthorblockN{
Wood Moore, Hayden Jananthan, Jeremy Kepner
\\
\IEEEauthorblockA{
MIT
}}}

\maketitle

\begin{abstract}
The PALU model is a hybrid power law generative framework
that decomposes Internet traffic graphs into three structural
components: a preferential attachment core, degree one leaf
nodes, and disconnected satellite clusters. Although the
original PALU formulation provided a theoretical basis for
this decomposition, it was never validated against real
traffic data and was restricted to unweighted networks. This
work presents the first empirical evaluation of PALU against
real Internet traffic, using packet traces from the MAWI
WIDE samplepoint F dataset. We introduce a weighted extension
of the model incorporating strength degree scaling, and
develop a packet-to-graph pipeline for parameter estimation
across multiple observation window sizes. Results show that
the power law degree exponent $\alpha \approx 2.0$ remains
stable across observation scales from $10^5$ to $3 \times
10^6$ packets, consistent with the PALU structural prediction.
However, the satellite cluster parameter $\lambda$ and the
weighted scaling parameter $\theta$ exhibit systematic
dependence on observation scale, and the Poisson star
assumption for unattached components is formally rejected
at all tested window sizes. These results indicate that
PALU captures important aspects of Internet traffic
structure but requires an explicit observation model to
fully account for finite sampling effects.
\end{abstract}

\section{Introduction}
\subsection{Motivation}

Internet traffic is fundamental to modern infrastructure, and understanding how Internet traffic flows is important for the overall management of the systems making up that infrastructure. Network traffic analysis can help inform capacity planning, network design, cybersecurity, and many other aspects of modern infrastructure. By treating devices as nodes and communication between those nodes as edges, Internet traffic can be represented as a graph, opening up Internet traffic to analysis by graph theory. This graph-based framing draws on a broader body of statistical physics work characterizing the structural properties of complex networks \cite{albert2002statistical}, including foundational results on small-world structure showing that many real networks combine high local clustering with short path lengths \cite{dj1998collective, watts1999small}. This allows for research to study the global structural properties, as well, as opposed to just analyzing individual packets. 
See \cite{newman2011structure} for a broader collection of foundational work in this area.


The scale of the Internet creates a huge modeling challenge due to the enormous size of these graphs. Additionally, these graphs often follow a heavy-tailed distribution, meaning these graphs are not easily explained by simple random network models. Network growth models, such as preferential attachment, provide one possible way to capture these structures.

\subsection{Preferential Attachment and Internet Traffic}

Barabási and Albert's preferential attachment \cite{barabasi1999emergence} showed that heavy-tailed distributions emerged naturally from a simple growth model. Later work showed that this power law scaling accurately encapsulates the topology of the Internet \cite{Albert_1999, yook2002modeling}. With empirical measurements of real Internet topology snapshots confirming strong power law relationships in metrics such as node out degree \cite{faloutsos1999power}, more recent large-scale measurement work has identified additional structural phenomena in Internet traffic that a pure power law model does not fully capture, motivating decomposed models such as PALU \cite{kepner2022new}. This body of work created the foundation for later Internet traffic modeling, including the model evaluated in this paper.


The PALU model is an extension of the preferential attachment model \cite{devlin2021hybrid}. This model includes three different structural components, the preferential attachment core, leaf nodes of one degree, and unattached satellite clusters. Through decomposing Internet traffic graphs into these elements, the model encapsulates a stronger statistical structure than a preferential attachment model would alone. 

The PALU model is characterized by several parameters describing both network topology and traffic intensity. $\alpha$ denotes the exponent of the power law degree distribution and $\lambda$ describes the size distribution of disconnected satellite clusters. Quantities $C$, $L$, and $U$ represent fractions of nodes belonging to core, leaf, and unattached components. The strength degree relationship is modeled as $s(k)\propto k^\beta$, with $\theta=\beta - 1$. 


Although the original PALU work \cite{devlin2021hybrid} created a mathematical understanding of this structure, it did not validate the model against real Internet traffic data. This research addresses an empirical gap rather than simply reproducing an existing experiment. Additionally, the original PALU framework is unweighted. This work attempts to extend the PALU model into a weighted extension incorporating strength degree scaling.






\section{Weighted PALU Model}

\subsection{PALU Network Structure}

The PALU model represents Internet traffic using a preferential attachment core, degree one leaf nodes, and unattached satellite components \cite{devlin2021hybrid}. The core ($C$) represents the primary connected portion of the networks where new connections are form preferentially. This produces the heavy tailed degree distribution of the model. Leaf nodes ($L$) are nodes connected to the core through a single edge and therefore have a degree of one. Unattached nodes ($U$) are smaller disconnected groups of nodes that don't interact with the main core.

\subsection{Unweighted PALU Formulation}

The PALU model characterizes the observed network using the total number of nodes $N_V$ and the three structural fractions $C$, $L$, and $U$. These fractions satisfy the constraint
\(
    C + L + U\left(1 + \lambda - e^{-\lambda}\right) = 1,
\)
where the unattached multiplier $\left(1 + \lambda - e^{-\lambda}\right)$ accounts for the expected number of nodes per unattached cluster under the Poisson satellite model. The total number of nodes in each component can therefore be represented as $N_C = CN_V$, $N_L = LN_V$, and $N_U = UN_V$.

The degree distribution of the preferential attachment core follows a power law form, $P(k) \propto k^{-\alpha}$, where $\alpha$ is the degree distribution exponent. The parameter $\lambda$ characterizes the size distribution of the unattached satellite components. This use of a Poisson-based stochastic model for growth processes within a preferential attachment framework parallels other extensions of the Barabási Albert model, such as Poisson growth preferential attachment, which similarly derives closed form likelihoods for parameter estimation directly from observed network snapshots \cite{sheridan2008preferential}.

\subsection{Weighted Extension}

The weighted extension of PALU incorporates the volume of communication
between nodes through edge weights. For a node $i$, the strength is defined as
$s_i = \sum_j w_{ij}$ where $w_{ij}$ is the weight of the edge between nodes
$i$ and $j$. This strength based framework builds on prior work establishing
that edge weights in real world networks are not randomly distributed across
topology, with high degree hubs tending to accumulate disproportionately large
total weight \cite{barrat2004architecture}. While the degree $k_i$ measures
the number of connections node $i$ has, its strength $s_i$ measures the total
weight of those connections.

The relationship between strength and degree can be modeled by $s(k) \propto k^\beta$ where $\beta$ describes how the traffic scales with connectivity. Therefore, the weighted PALU parameter $\theta$  is defined as $\theta = \beta - 1$.

Therefore, $\theta$ captures the relationship between a node's connectivity and the traffic of it's connections. For this work, $\beta$ is estimated from empirical strength degree relationship and $\theta$ is calculated as a result.

\subsection{Observation/Subsampling Model}

The MAWI WIDE traces only provide a finite observation window of Internet traffic and therefore cannot capture the complete underlying communication network. The observed graph can only be viewed as a sampling of the underlying network, $G_{\text{observed}} \sim \text{Sample}(G_{\text{underlying}}, p)$, where $p$ represents the probability that an underlying communication edge is observed within a measurement window. This sampling process is a critical methodological concern, prior work has shown that even random graphs with no true scale free structure can appear to exhibit a power law degree distribution when sampled via a limited number of vantage points, since the sampling process itself can introduce a heavy tailed artifact \cite{clauset2003traceroute}. Though, as the observation window increases, more communication events could be captured, which could potentially change the observed network structure.

\section{Dataset and Experimental Methodology}


The MAWI Working Group WIDE Project (MAWI WIDE) provides packet traces from observed Internet network traffic. This work uses traffic from Samplepoint F, which provides daily packet level measurements. Each trace contains a 15 minute observation window. They contain anonymized network addresses.

For this work, the packet traces were used to construct communication graphs from observed traffic. The relevant packet fields are the source IP address, destination IP address, and packet size in bytes. Source and destination addresses identify which nodes are communicating, while packet size determines the weight of each communication edge. This structure allows for packet data to be represented as a graph for PALU analysis. 


The MAWI packet traces are processed through a series of steps to construct the graph used for PALU analysis. 



From this resulting graph, the degree and strength of each node are calculated. The graph is then decomposed into the PALU structural components and the model parameters are estimated for each window. This pipeline allows the MAWI packets traces to be analyzed consistently across different traffic windows. 



MAWI traffic is inherently directed, since communication from a source IP to a destination IP is distinct from communication in the opposite direction. However, the PALU representation used in this study treats the network as undirected. As a result, traffic between two IP addresses is represented as a single connection rather than separate source to destination and destination to source edges. 





Analysis was performed using multiple observation window sizes, i.e., $N_V \in \{10^5, 3\times10^5, 10^6, 3\times10^6\}$, allowing the effect of the sampling scale on the observed network to be evaluated. The main prediction is that the PALU parameters should remain stable across windows, while the observed network itself changed as more traffic is captured.


To reduce potential sampling bias, the preprocessing pipeline randomly shuffles the packets before construction of observation windows. Various randomized windows are generated at each window size and the PALU parameters are estimated for each, providing multiple samples at each observation scale and allowing parameter stability to be evaluated without a single fixed subset of packets.

\section{Parameter Estimation}


The power law tail is estimated using the degrees of nodes belonging to a PALU core, extracted from an observed network. Then the lower bound of the power law region, $k_{\min}$, and the value producing the smallest Kolmogorov Smirnov (KS) statistic is selected. The exponent $\alpha$ is estimated using a discrete Clauset-Shalizi-Newman-style maximum likelihood estimator \cite{Clauset_2009}. The resulting KS statistic is reported as a measure of the agreement between the empirical tail and the fitted power law model.


For each disconnected component, the satellite nodes are identified and their sizes recorded. A zero truncated Poisson distribution is fitted to the different observed satellite sizes and the parameter $\lambda$ estimated from the result.


The structural composition of each observation window is quantified into its fractions of nodes belong to the core, leaf, and unattached components.
\begin{equation*}
    C = N_{\mathrm{core}}/N, \quad L = N_{\mathrm{leaf}}/{N}, \quad U = N_{\mathrm{unattached}}/{N},
\end{equation*}
where $N$ is the total number of observed nodes. These fractions describe the distribution of the networks nodes across the PALU components. 

These structural fractions are evaluated across the different window observation sizes. As the observation window increases, additional communication is captured and the observed network structure can change. The core fraction, in particular, os expected to increases as more nodes become connected to the core. Comparing $C$, $L$, and $U$ across window sizes provides another measure of how the PALU structure changes with sampling.

\subsection{Strength Degree Scaling}

To evaluate the weighted extension of PALU, the relationship between node strength and degree is modeled as 
\(
    s(k) \propto k^{\beta},
\)
where $s(k)$ is the average strength of nodes with degree $k$, and $\beta$ describes how traffic volume scales with
connectivity. The weighted parameter is calculated as $\theta = \beta - 1$.
The exponent $\beta$ is estimated via ordinary least squares regression on $\log s$ against $\log k$. The goodness of fit is reported as the coefficient of determination $R^2$. At small observation scales ($N_V = 10^5$), the fitted $R^2 \approx 0.18$, indicating that the strength degree relationship is poorly constrained at small window sizes due to an insufficient number of high-degree nodes. Estimates of $\theta$ at small $N_V$ should therefore be interpreted with caution. At larger observation scales ($N_V = 3 \times 10^6$), $R^2$ increases to approximately $0.45$, providing more reliable estimates. The strength degree relationship provides the principal test of the proposed weighted extension.

\section{Experimental Design}

\subsection{Experiment 1 -- Power Law Validation}

The first experiments test whether MAWI traffic exhibits predicted heavy tailed degree distribution. For each given observation window, the empirical degree distribution of the PALU core is compared with its fitted power law distribution. The analysis reports the estimated exponent $\alpha$, the selected $k_{\min}$, and the corresponding KS statistic to quantify overlap between empirical and fitted distributions. 

Evidence supporting the hypothesis that MAWI degree distributions exhibit a PALU-consistent power law tail would be a clear heavy tailed degree distribution with a power law tail and a relatively small KS statistic for the fitted region.

\subsection{Experiment 2 -- Cross Window Parameter Stability}

The primary experiment evaluates whether the PALU parameters remain stable as the amount of observed traffic increases. For each observation window size, $N_V \in \{10^5,3\times10^5,10^6,3\times10^6,\ldots\}$,
the parameters $\alpha,\lambda,\theta,C,L,$ and $U$ are independently estimated. The resulting estimates are then plotted as a function of $N_{V}$. 

The hypothesis is that, although the observed network structure changes with increasing observation size, the PALU parameters should remain approximately stable. Therefore, consistent parameter estimates across window sizes would provide evidence that the model captures structural properties of the traffic rather than sampling effects of a particular observation scale.

Parameter estimates can be analyzed as a function of $\log N_{V}$. Regression slopes and confidence intervals can be used to evaluate systematic trends, while coefficients of variation can quantify relative variability across window sizes. If repeated windows are available, bootstrap confidence intervals or group comparison tests such as ANOVA or Kruskal Wallis can be used to assess whether the parameter estimates differ significantly across observation window sizes.

\section{Results}

\subsection{Parameter Stability}

The PALU parameters were estimated independently for each observation window. Tables~\ref{tab:palu_params_top},\ref{tab:palu_params_bot} summarize the mean and standard deviation of each parameter across the three windows at each observation scale.

\begin{table}[h]
\centering
\caption{Mean $\pm$ standard deviation of estimated PALU
structural parameters across three windows at each observation
scale.}
\label{tab:palu_params_top}
\begin{tabular}{lcccc}
\hline
$N_V$ & $\alpha$ & $\lambda$ & $C$ & $L$ \\
\hline
$10^5$ &
    $1.89 \pm 0.04$ &
    $2.59 \pm 0.14$ &
    $0.09 \pm 0.003$ &
    $0.48 \pm 0.02$ \\
$3 \times 10^5$ &
    $1.96 \pm 0.07$ &
    $2.23 \pm 0.08$ &
    $0.18 \pm 0.03$ &
    $0.61 \pm 0.04$ \\
$10^6$ &
    $1.99 \pm 0.08$ &
    $1.75 \pm 0.02$ &
    $0.29 \pm 0.01$ &
    $0.63 \pm 0.005$ \\
$3 \times 10^6$ &
    $1.99 \pm 0.05$ &
    $1.70 \pm 0.03$ &
    $0.55 \pm 0.02$ &
    $0.43 \pm 0.02$ \\
\hline
\end{tabular}
\end{table}

\begin{table}[h]
\centering
\caption{Mean $\pm$ standard deviation of estimated PALU
weighted and unattached parameters across three windows at
each observation scale.}
\label{tab:palu_params_bot}
\begin{tabular}{lccc}
\hline
$N_V$ & $\theta$ & $U$ & $R^2$ \\
\hline
$10^5$ &
    $0.05 \pm 0.04$ &
    $0.43 \pm 0.03$ &
    $0.18 \pm 0.01$ \\
$3 \times 10^5$ &
    $0.15 \pm 0.10$ &
    $0.21 \pm 0.02$ &
    $0.25 \pm 0.01$ \\
$10^6$ &
    $0.31 \pm 0.13$ &
    $0.07 \pm 0.01$ &
    $0.37 \pm 0.04$ \\
$3 \times 10^6$ &
    $0.38 \pm 0.10$ &
    $0.02 \pm 0.003$ &
    $0.43 \pm 0.03$ \\
\hline
\end{tabular}
\end{table}

The estimated $\alpha$ remains in the range $1.85$--$2.05$ across all window sizes, consistent with the power law degree structure predicted by the PALU model. In contrast, $\lambda$ decreases systematically from approximately $2.6$ at $N_V = 10^5$ to approximately $1.7$ at $N_V = 10^6$, where it stabilizes. The weighted parameter $\theta$ increases monotonically from near zero at small windows to approximately $0.38$ at $N_V = 3 \times 10^6$, with notably low goodness of fit ($R^2 \approx 0.18$) at small observation scales, indicating that $\theta$ estimates at small $N_V$ are unreliable rather than structurally meaningful. The structural fractions $C$ and $U$ show strong and opposing trends with $N_V$.

\begin{figure*}[h]
    \centering
    \includegraphics[width=1.75\columnwidth]{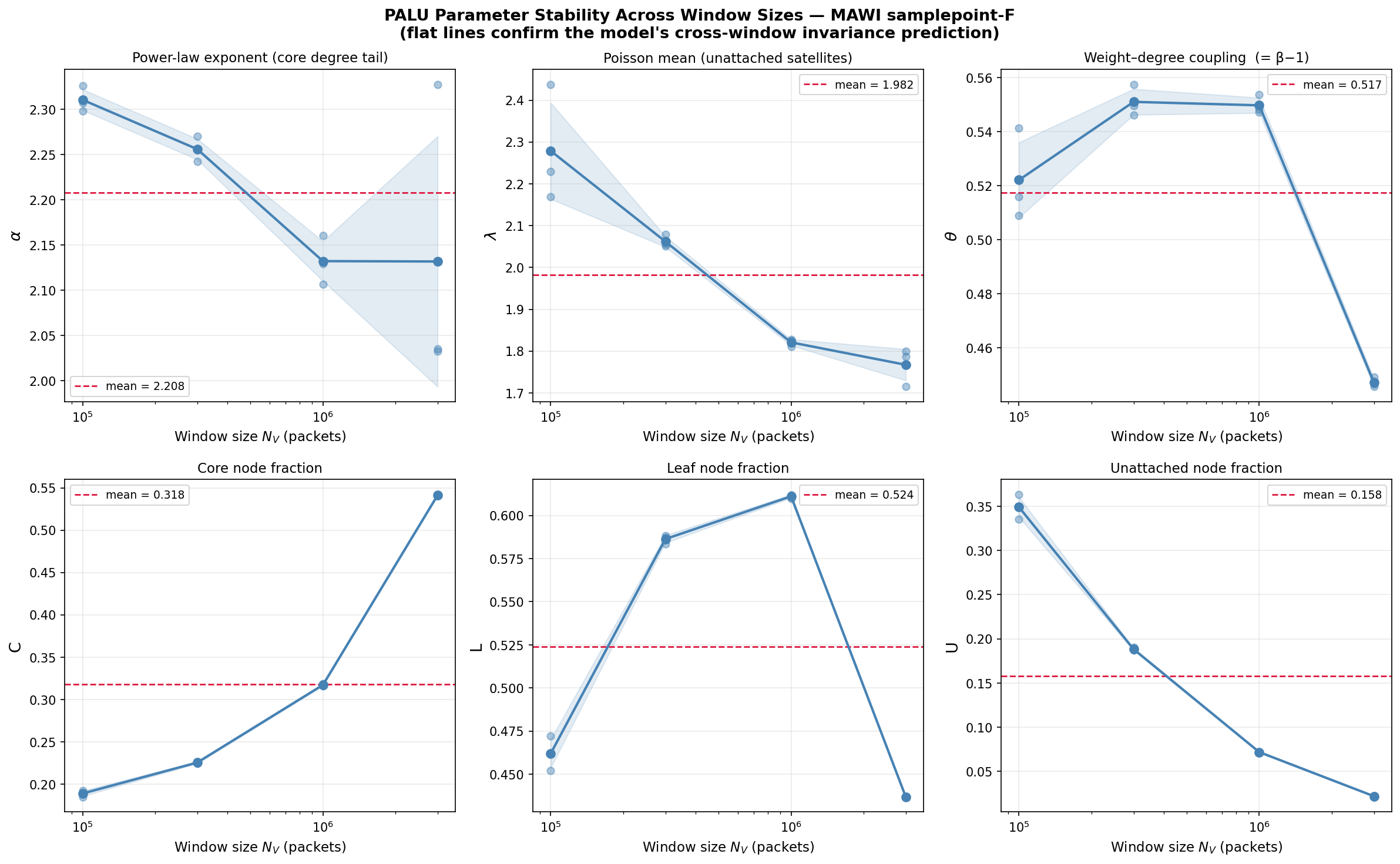}
    \caption{Variation of the PALU parameters with observation window size $N_V$. The estimated $\alpha$ remains relatively stable, whereas $\lambda$ and $\theta$ exhibit stronger dependence on observation size. The structural fractions show an increasing core fraction $C$, decreasing unattached fraction $U$, and a leaf fraction $L$ that peaks at intermediate observation scales before declining at larger window sizes.}
    \label{fig:stability}
\end{figure*}

\section{Discussion}


Among the measured parameters, the stability of $\alpha$ provides the strongest evidence supporting the PALU prediction for degree distribution. This stability is consistent with theoretical work showing that power law degree scaling persists under preferential attachment even when node heterogeneity is introduced into the attachment mechanism \cite{santiago2008extended}. However, because sampling procedures can themselves induce apparent power law behavior in networks that are not truly scale free \cite{clauset2003traceroute}, the stability of $\alpha$ alone should not be treated as conclusive evidence for the PALU mechanism without also ruling out sampling induced artifacts.

The results also show increasing connectivity as more traffic is observed, with a meaningful preferential attachment like structure. Among the measured parameters, the stability of $\alpha$ provides the strongest evidence supporting the PALU prediction for degree distribution. 


The estimated $\lambda$ shows a strong dependence on observation window size. Several factors may contribute to this, including finite sampling, dominance of two node components at smaller windows, persistent scanner traffic, component fragmentation, and the assumption that unattached components can be represented by the star-like structure used. 

The weighted parameter $\theta$ exhibits strong scale dependence, particularly important for evaluating the proposed weighted extension. Under the current observation model, the assumption that a single $\theta$ characterizes the underlying weighted network may not hold across different observation window sizes. This scale dependence provides a scientifically meaningful result about the sensitivity weighted network structure to observation scale. 


As observation window sizes $N_{V}$ increases, core fraction increases while unattached fraction decreases, which can be explained by the increased amount of communication captured in larger windows. Additional observed edges can connect nodes that appeared isolated or disconnected in smaller windows to the main component. As these previously unattached nodes become connected to the larger network, they contribute to the core or connected structure, causing $C$ to increase and $U$ to decrease.







\section{Limitations}

There are several limitations to this study:
\begin{itemize}
    \item Real Internet traffic is inherently directed, but the PALU representation assumes an undirected network.
    \item The PALU model characterizes network structure entirely through degree-based decomposition into 'core', 'leaf', and 'unattached' components, but doesn't account for other well-established structural properties of real world networks, e.g., high local clustering combined with short characteristic path lengths  \cite{dj1998collective}. Two networks with identical PALU parameters ($\alpha$, $\lambda$, $C$, $L$, $U$) could still differ substantially in their clustering or path length structure.
    \item The observed graph is not necessarily the complete underlying communication network. 
    \item Persistent ICMP scanning traffic may influence the unattached component and affect the estimation of $\lambda$.
    \item The zero truncated Poisson fit for $\lambda$ was formally rejected by the Kolmogorov Smirnov test at every observation window size tested, with $p \approx 0$ in all cases, indicating the satellite cluster sizes in observed MAWI traffic do not follow the assumed Poisson star distribution. 
    \item The PALU model, like the underlying preferential attachment framework, assumes a purely growing network, but MAWI traffic does not grow monotonically, nodes and connections continuously appear and disappear across the observation window. Recent work shows introducing node removal into a preferential attachment process can produce a topological phase transition, shifting the degree distribution from a power law to one with an exponential tail, depending on the relative rates of removal and attachment \cite{bauke2011topological}. This suggests that the heavy tailed degree structure observed should not be assumed to hold unconditionally; node churn within the observation window may be an additional factor influencing the stability of the estimated PALU parameters.
    \item At small observation sizes, small disconnected components, particularly two node components, can dominate the estimate of $\lambda$. This may introduce sampling effects that should be considered when interpreting the results.
    \item The current evidence for $\theta$ is weaker than the evidence for $\alpha$, suggesting a single value of $\theta$ may not adequately characterize the weighted network across observation scales. Furthermore, the low $R^2 \approx 0.18$ at small observation windows indicates that $\theta$ estimates at small $N_V$ reflect estimation noise rather than a true structural signal.
    \item This work assumes that the PALU core degree distribution is a strict power law; recent work has shown that many empirically observed degree distributions are better described by extensions or alternatives to the pure power law \cite{lee2024degree}. Although the KS statistic based fitting procedure used here provides some indication of goodness of fit, future work should consider whether alternative tail models describe the MAWI core degree distribution more accurately than the power law assumed by PALU.
\end{itemize}

\section{Conclusion}

We developed a weighted PALU model and a MAWI WIDE empirical validation pipeline. MAWI traffic exhibits a stable heavy tailed degree structure consistent with PALU, with $\alpha$ remaining relatively stable across observation scales. However, $\lambda$ and $\theta$ show systematic scale dependence. Thus, PALU captures important aspects of Internet traffic structure but does not fully explain the observed network under finite sampling. Future work should incorporate an explicit observation model, test additional MAWI traces, and compare PALU with alternative network growth models.

\section*{Acknowledgment}
The authors wish to acknowledge the following individuals for their contributions and support 
L. Anderson, W. Arcand, D. Bestor, W. Bergeron, B. Bond, C. Byun, A. Bonn, D. Burrill, V. Gadepally, J. Gottschalk, T. Hardjono, M. Houle, M. Hubbell, M. Jones, P. Luszczek, P. Michaleas, L. Milechin, J. Mullen, C. Leiserson, C. Milner, S. Mohindra,  A. Pentland, A. Prout, C. Prothmann, A. Reuther, A. Rosa, D. Rus, D. Ross, J. Ross, M. Sherman, S. Van Broekhoven, M. Weems, J. Wilkinson, C. Yee.

\bibliographystyle{IEEEtran}
\bibliography{references}

\end{document}